\documentclass[12pt,preprint]{aastex}
\usepackage{url}

\begin{document}

\title{The Total Errors In Measuring $E_{peak}$ for Gamma-Ray Bursts}
\author{Andrew C. Collazzi, Bradley E. Schaefer $\&$ Jeremy A. Moree \affil{Physics and Astronomy, Louisiana State University, Baton Rouge, LA 70803}}

\begin{abstract}
Of all the observable quantities for Gamma-Ray Bursts, one of the most important is $E_{peak}$. $E_{peak}$ is defined as the peak of the $\nu F_\nu$ power spectrum from the prompt emission. While $E_{peak}$ has been extensively used in the past, for example with luminosity indicators, it has not been thoroughly examined for possible sources of scatter. In the literature, the reported error bars for $E_{peak}$ are the simple Poisson statistical errors.  Additional uncertainties arise due to the choices made by analysts in determining $E_{peak}$ (e.g., the start and stop times of integration), imperfect knowledge of the response of the detector, different energy ranges for various detectors, and differences in models used to fit the spectra.  We examine the size of these individual sources of scatter by comparing many independent pairs of published $E_{peak}$ values for the same bursts.  Indeed, the observed scatter in multiple reports of the same burst (often with the same data) is greatly larger than the published statistical error bars.  We measure that the one-sigma uncertainty associated with the analyst's choices is 28\%, i.e., 0.12 in $Log_{10}(E_{peak})$, with the resultant errors always being present.  The errors associated with the detector response are negligibly small.  The variations caused by commonly-used alternative definitions of $E_{peak}$ (such as present in all papers and in all compiled burst lists) is typically 23\%-46\%, although this varies substantially with the application.  The implications of this are: (1) Even the very best measured $E_{peak}$ values will have systematic uncertainties of 28\%.  (2) Thus, GRBs have a limitation in accuracy for a single event, with this being reducible by averaging many bursts.  (3) The typical one-sigma total uncertainty for collections of bursts is 55\%.  (4) We also find that the width of the distribution for $E_{peak}$ in the burst frame must be near zero, implying that some mechanism must exist to thermostat GRBs. (5) Our community can only improve on this situation by using collections of bursts which all have identical definitions for the $E_{peak}$ calculation.
\end{abstract}


\section{Introduction}

$E_{peak}$, the peak of the $\nu F_\nu$ power spectrum from the prompt emission of a long-duration Gamma-Ray Burst (GRB), is one of the most important quantities measured from a GRB.  With GRB spectra being essentially smoothly broken power laws with no sharp features (Band et al. 1993), the $E_{peak}$ value is the primary description of the entire spectrum.  Observed $E_{peak}$ values typically range from a few keV to over a few MeV (e.g. Kippen et al. 2002; Barraud et al. 2003; Schaefer 2003; Sakamoto et al. 2005; Sakamoto et al. 2008). This distribution is single-peaked (from 20-2000 keV) and fairly narrow (Mallozzi et al. 1995). It is unclear how X-Ray Flashes (XRFs) fit into this distribution. Two good examples of XRFs contribution to the distribution can be seen in Figure 7 of Sakamoto et al. (2005) and Figure 4 of P\'{e}langeon et al. (2008). In both these figures, there is a small marginally-significant secondary peak composed of XRFs. It is not yet clear whether this is a separate peak or merely an extended tail from the originally found GRB distribution (as seen in Mallozzi et al. 1995).

A decade ago, two separate groups made the discovery that easily measurable properties are well correlated with the peak luminosity of a GRB (Norris et al. 2000; Fenimore $\&$ Ramirez-Ruiz 2000). Soon thereafter, three correlations were identified between $E_{peak}$ and either the burst peak luminosity or the total energy (Amati et al. 2002; Schaefer 2003; Ghirlanda et al. 2004). With these luminosity relations between measurable properties and the luminosity, GRBs became `standard-candles.'  That is, just like Cepheids and Type Ia supernovae, we can measure light curve or spectral properties, use the luminosity relations to estimate the luminosity, and then use the observed brightness and the inverse-square law of light to derive the GRB distance.  This means that the distances of GRBs can be determined without relying on spectroscopic redshifts, which offers a means for estimating the luminosity and hence redshift for the $\sim70\%$ of bursts with no measured spectroscopic redshift.  In comparison with spectroscopic redshifts, the GRB luminosity relations have the big disadvantage of providing relatively poor accuracy, but they have the big advantages of providing unbiased redshifts for almost all bursts for demographic purposes (Xiao \& Schaefer 2010) and of providing independent luminosity distances for Hubble Diagram purposes (Schaefer 2007).

In all, seven luminosity relations have been discovered and confirmed through additional testing (Schaefer 2007; Schaefer \& Collazzi 2007).  The most publicized of these relations is the Amati relation, which identifies the total burst gamma-ray energy (assuming isotropic emission) as a power law of $E_{peak}$, corrected for redshifting to the rest frame of the burst (Amati et al. 2002; Amati 2006). Despite the Amati relation getting the most attention, the so-called `Ghirlanda relation'  (Ghirlanda et al. 2004) has the best accuracy (as measured by the amount of scatter in the calibration curve). It connects the total burst energy (corrected for the anisotropic emission of the relativistic jet) as a power law to $E_{peak}$. The third luminosity relation relates $E_{peak}$ with the burst peak luminosity, L (Schaefer 2003; Yonetoku et al. 2004). Other relations involve correlating L with the spectral lag (Norris et al. 2000), the rise-time (Schaefer 2002), variability (Fenimore $\&$ Ramirez-Ruiz 2000), and the number of peaks in the light curve (Schaefer 2002). These seven relations have been confirmed by many different groups demonstrating the same relation in independent samples of bursts (e.g. Schaefer et al. 2001; Reichart et al. 2001;  Amati 2003;  Bloom et al. 2003; Amati 2006; Li $\&$ Paczynski 2006; Nava et al. 2006; Schaefer 2007; Butler et al. 2009).

A variety of different problems have been raised regarding the luminosity relations, many of which focus on one specific relation or another. For example, the Amati relation has an ambiguity when the measured properties are used to determine the redshift (Li 2006; Schaefer $\&$ Collazzi 2007), the Ghirlanda relation can only be applied to the small fraction of bursts with a known jet break, the identification of jet breaks has become confused when the X-ray afterglow light curves are considered (Melandri et al. 2008), the `variability' relation suffers from issues tied to how variability is defined (Schaefer 2007), and the number-of-peaks relation only provides a limit on the luminosity (Schaefer 2007). Another proposed luminosity relation (Firmani et al. 2006) has been shown to provide no improvement upon previously existing ones, and indeed can be directly derived from the prior luminosity relations (Collazzi $\&$ Schaefer 2008). In addition, a variety of new luminosity relations have been proposed and have yet to be extensively tested (e.g. Dainotti et al. 2008).  These various problems can be well handled, mainly by the careful use of the relations and their input.

By far the greatest problem with all the luminosity relations is accuracy. The most accurate of the luminosity relations (the Ghirlanda relation) has an RMS scatter about its calibration line of 0.15 in the log of the luminosity.  Meanwhile, the weakest of the luminosity relations (the variability and rise-time relations) have an RMS scatter about their calibration lines of 0.45 in the log of their luminosity. When the resultant luminosities for the relations for a single burst are combined as a weighted average, the average uncertainty is 0.26 in log-luminosity (Schaefer 2007).  This translates into an average one-sigma error in distance modulus ($\sigma_\mu$) of 0.65 mag.  This error is greatly larger than those from  optical spectroscopy, yet this poorer accuracy is fine for many GRB demographic studies.  For Hubble diagram work, our community will compare the $\sigma_\mu=0.65$ mag accuracy for GRBs with those of the Type Ia supernovae.  For comparison, supernovae have $\sigma_\mu=0.36$ mag (Perlmutter et al. 1999), $\sigma_\mu=0.29$ mag even after heavy selection to create the `gold sample' (Riess et al. 2004) and $\sigma_\mu>0.25$ mag from the Supernova Legacy Survey (Astier et al. 2006).  For some sort of an average of $\sigma_\mu \approx 0.30$ mag for supernovae, we see that a single GRB has an accuracy that is $2.1\times$ worse than that of a single supernova.  This is much better than some people might expect.  For Hubble diagram work, GRBs provide unique information on the expansion history of the Universe for redshifts from 1.7 to 8.2. 

A primary task for the GRB community is to substantially improve the accuracy of the luminosity relations.  Some of the scatter in the current calibration might be caused by apparently random fluctuations in the source resulting in variations of the burst luminosity even for bursts with identical measured indicators. Another source of scatter might be that we simply cannot (or have not) measured the luminosities and indicators with sufficient accuracy. That is, the scatter in the luminosity relations might owe part of its scatter to systematic uncertainties in the luminosity indicators. However, it is not entirely clear how scatter in $E_{peak}$ will effect the scatter in the associated luminosity relations. This is largely because a error in finding $E_{peak}$ will also result in a mis-calculation of the factors used in the associated luminosities, $E_{\gamma}$, L and $E{\gamma,iso}$. Therefore, it is difficult to quantify just how much the scatter in finding $E_{peak}$ will scatter the luminosity relations. Nonetheless, it is clear that understanding just how much scatter there is in the measurement of $E_{peak}$ is important to work on luminosity relations.

Our group has been closely evaluating and optimizing the various luminosity relations (e.g., Schaefer \& Collazzi 2007; Collazzi \& Schaefer 2008, Xiao \& Schaefer 2009), so we have started a program to evaluate the real total uncertainties in the various luminosity relations.  The $E_{peak}$ quantity is the most prominent luminosity indicator (and of high importance for many other applications), so we began with this.  We first examine the sources of uncertainty that arise in measuring $E_{peak}$.  We go beyond the usual measurement errors derived from Poisson statistics as reported in all papers, and we look at various sources of systematic errors.  We then quantify these sources of uncertainty, with our primary tool being the comparison of multiple independent published values of $E_{peak}$ reported for the same bursts.

\section{Types of Uncertainty in $E_{peak}$}

When a burst occurs, there are a variety of ways in which uncertainty is added. The most familiar source of uncertainty is the ordinary Poisson variations in the number of photons that appear in each energy bin, resulting in random variations in the measured $E_{peak}$. This statistical error ($\sigma_{Poisson}$ ) is what is reported in the literature when values of $E_{peak}$ are given. 

A second issue that arises in determining $E_{peak}$ is the various choices that are made by the analyst. These choices include the exact time and form of the background light curve, the exact time interval over which to accumulate the spectrum, the energy range for the spectral analysis (which is often smaller than the full range of the instrument), and even the convergence criteria for the fit.   Identical burst data can be fit by two independent analysts with two entirely different (yet reasonable) sets of choices, resulting in different $E_{peak}$ values.  Neither of these values can be identified as being right or wrong, nor can we know which one is better. Therefore, this difference between the two is a type of uncertainty, $\sigma_{Choice}$. 

A third source of uncertainty comes from not knowing the detector response perfectly, which can be characterized as imperfect calibration of the detector response matrix.  We will call the resultant errors on $E_{peak}$ $\sigma_{Det}$. Another component of $\sigma_{Det}$ is the energy range of different detectors. Two satellites can yield different values for $E_{peak}$ merely as a result of covering different energy ranges. This would occur when one satellite gets a better profile of the `turnover' of the spectral profile than another. 

The final source of uncertainty is related to the specific definition of $E_{peak}$.  While, at first glance, the $E_{peak}$ has a simple definition, there are actually a variety of alternatives that are commonly used.  Each of these definitions produces a different value, and this appears as a systematic uncertainty, which we label $\sigma_{Def}$.  We can point to four alternative definitions:  (1) The GRB spectrum can be fit either to the Band model, a smoothly broken power law (Band et al. 1993) or to the `Comptonized Power Law' model, a power law times an exponential cutoff.  (2) The GRB spectrum can be extracted for the entire burst (a `fluence spectrum') or for just the time of the peak flux.  The fluence spectrum is relevant to the Amati and Ghirlanda relations (which use the burst fluence), while the peak flux is relevant for the other relations (which use the burst peak luminosity).  Problems with the use of the peak spectrum are that the number of photons are usually low (leading to poor accuracy) and that the time range for extracting the spectrum is not defined (leading to variations due to the choice of interval).  $E_{peak}$ varies substantially throughout most bursts (e.g. Ford et al. 1995), so the choice of the time interval makes for large uncertainty.  (3) The high -energy and low-energy power law indices for the Band function can either be fitted to the spectrum or they can be set to average values.  When the spectrum does not extend much above $E_{peak}$, many analysts will simply set the high-energy index equal to some average.  This common practice leads to systematically different $E_{peak}$ values.  (4) The analyst might define the $E_{peak}$ value based on traditional frequentist method, or they might impose various priors within a Bayesian method.  Depending on the adopted priors, the Bayesian method can give greatly different values than the frequentist method.

The luminosity relations are all expressed as power laws, which is appropriate for the physical derivations of the relations, and the various errors are multiplicative. Therefore it is best to consider the logarithm of the relevant quantities, for example, Log($E_{peak}$). The \textit{total} measurement uncertainty of Log($E_{peak}$) will be labeled as $\sigma_{Total}$. Therefore, as the individual errors are additive on a logarithmic scale, the total error will be:
\begin{equation}
\label{eq:sigtot}
\sigma^2_{Total} = \sigma^2_{Poisson} + \sigma^2_{Det} + \sigma^2_{Choice} +\sigma^2_{Def}.
\end{equation}
Our task is now to derive $\sigma_{Total}$ by determining the remaining three sources of individual errors (as $\sigma_{Poisson}$ is already reported in the literature).

Our general procedure for isolating the various sources of errors will be to compare two measured $E_{peak}$ values, $E_{peak,1}$ and $E_{peak,2}$, that have identical conditions except for some difference.  We quantify this difference as 
\begin{equation}
\label{eq:delta}
\Delta = Log_{10}(E_{peak,1}) - Log_{10}(E_{peak,2}).
\end{equation}
In general, we will be evaluating $\Delta$ for various sets of bursts, for example with the values from one source all being denoted with the subscript `1' and some other source being denoted with the subscript `2'.  With many measures of $\Delta$, the average will generally be near zero and there will be some RMS scatter, denoted as $\sigma_{\Delta}$.  The scatter of the $\Delta$ values will be a measure of the uncertainty arising from the differences in the input.

\section{Specific Examples}

The essence of the problem and of our method comes from a comparison of $E_{peak}$ values as reported for many different satellites, analysts, and models.  In the following sections, our analysis will highlight abstract statistics for which it is easy to lose the real picture that the published $E_{peak}$ values have much larger scatter than we would expect from systematic errors alone.  Below, we provide four specific examples of GRBs. In some cases, e.g. Butler et al. (2007), we had to convert the reported error bars from their stated 90\% confidence values into their standard one-sigma values. Therefore, all uncertainties below are at the one-sigma level.

GRB 910503 (BATSE trigger 143) was one of the brightest bursts seen by BATSE.  Independent reports on $E_{peak}$ give $466\pm4$ keV (Band et al. 1993), $741$ keV (Schaefer et al. 1994), $621\pm11$ (Yonetoku et al. 2004), and $586\pm28$ (Kaneko et al. 2006).  All of these values have small statistical error bars, and all are separated from each other by much more than these error bars.  All these measures use identical data and models, so the wide divergence must be due to specific choices made by the individual analyst.  The $Log_{10}(E_{peak})$ values are $2.668\pm0.004$, $2.87$, $2.793\pm0.008$, and $2.768\pm0.021$.  The RMS scatter is 0.083 (which is greatly larger than all the $\sigma_{Poisson}$ values), which should equal to $\sigma_{Choice}$ for this one burst.

GRB 911109 (BATSE trigger 1025) is a burst near the BATSE median brightness level for which we found five independent measures of $E_{peak}$.  Band et al. (1993) give $114\pm 3$ keV, Schaefer et al. (1994) give $125$ keV, Yonetoku et al. (2004) give $153.2^{+7.5}_{-7.1}$ keV, Kaneko et al. (2006) give $131\pm6$ keV, and Nava et al. (2008) give $117\pm74$ keV.  Again, we see a scatter greatly larger than the quoted error bars.  For this one burst, the RMS scatter gives $\sigma_{Choice}=0.05$.

GRB 050525A was a very bright burst detected by four instruments.  Swift data gives $78.8^{+2.4}_{-1.8}$ keV (Blustin et al. 2006), $82^{+2.4}_{-1.8}$ (Sakamoto et al. 2008), $82^{+2.4}_{-1.8}$ keV (Butler et al. 2007), $81\pm3$ keV with a Bayesian analysis (Butler et al. 2007), and $102.4^{+4.8}_{-4.0}$ keV for a time interval including only the peak of the burst (Blustin et al. 2006).  The first three of these values from \textit{Swift} are found using identical models and data, so the variations can only arise from analyst choices, which for a very bright burst will have relatively small effect on the spectrum.  (In particular, it does not really matter what the choices for the background fit are because the background is so small compared to the burst flux.  Also, with a very bright burst, the start and stop times are well defined so that analyst choices will be very close.)  Nevertheless, two separate analyses of the identical data from INTEGRAL IBIS data gives either $69\pm72$ (Foley et al. 2008) or $58_{-21}^{+29}$ keV (Vianello et al. (2009), with these values not being so close.  For measures with other instruments, INTEGRAL SPI data gives $80\pm28$ keV (Foley et al. 2008), and Konus-Wind data gives $84.1\pm 1.7$ keV (Golenetskii et al. 2005a).

GRB 070508 was a bright burst detected by four satellites.  Konus-Wind data gives $188\pm5$ keV (Golenetskii et al. 2007), Suzaku data gives $233\pm7$ keV (Uehara et al. 2007), and RHESSI data gives $254^{+43}_{-27}$ keV (Bellm et al. 2007).  These values are inconsistent with any constant, implying that there must be additional systematic uncertainties past the reported statistical error bars.  $E_{peak}$ values have also been reported many times for Swift data, with the first circular giving $258\pm80$ keV (Barthelmy et al. 2007), the Sakamoto et al. (2008) catalog giving $260^{+122}_{-41}$ keV, an independent analysis giving $210^{+48}_{-24}$ keV (Butler et al. 2007), a Bayesian analysis giving $208^{+46}_{-25}$ keV (Butler et al. 2007), while a joint fit of the Swift-plus-Suzaku data gives $235\pm12$ keV for the Band function or $238\pm11$ for the CPL (Comptonized Power Law) function (Krimm et al. 2009).  The first four Swift values all use identical data and models, yet still the uncertainty for this bright burst runs from 210-260 keV.  Looking at all the reports, if we had to `vote on the truth' with an average, we would guess $E_{peak}\sim 230$, with this being dominated by the three `votes' controlled by the Suzaku data.  For all nine published values, a realistic analysis could take the $E_{peak}$ to be anywhere from roughly 190 to 260 keV.  And this is for a {\it bright} burst where all the problems are minimized.

\section{Typical $\sigma_{Poisson}$}

Ordinary Poisson fluctuations of the counts in each spectral energy bin result in an apparently random noise, which will somewhat shift the fitted $E_{peak}$ value.  This statistical uncertainty can be reliably calculated by keeping track of the counts and applying Poisson statistics, with the resulting uncertainties confidently propagated.  Most of the reported $E_{peak}$ values in the literature have reported error bars, and these are always from Poisson statistics alone.  These reported error bars are cast into log-base-10 and are labeled $\sigma_{Poisson}$.

The Poisson errors change greatly from burst to burst.  At one extreme for a very bright burst, GRB050525A has $E_{peak}=82^{+2.4}_{-1.8}$ keV (Sakamoto et al. 2008), with this being converted to $Log_{10}(E_{peak})=1.91\pm 0.01$.  At the other extreme are faint bursts with only poor constraints, for example BATSE trigger 658 with $E_{peak}=70\pm56$ keV (Nava et al. 2008), with this being converted to $Log_{10}(E_{peak})=1.85\pm 0.35$.  Throughout this paper, we will be using error bars on the log-base-10 of $E_{peak}$, where $\pm0.01$ corresponds to a 2.3\% error in $E_{peak}$, $\pm0.10$ corresponds to a 23\% error, and $\pm0.30$ corresponds to a factor of two error.

Collections of bursts with a wide range of individual error bars will have a much more restricted range of average error bars.  From 306 BATSE bursts, Kaneko et al. (2006) have error bars with average $\sigma_{Poisson}=0.04$.  From 37 HETE bursts, Sakamoto et al. (2005) have the average $\sigma_{Poisson}=0.17$.  From 9 INTEGRAL bursts (after excluding two with very large quoted error bars), Foley et al. (2008) have the average $\sigma_{Poisson}=0.35$.  From 32 Swift bursts, Sakamoto et al. (2008) have the average $\sigma_{Poisson}=0.08$.

An annoying problem is that recently some satellite programs have taken to reporting 90\% error bars rather than the universal standard one-sigma error bars.  This creates a problem when comparing the error bars with standard results or in doing any sort of statistical analyses.  The general solution is to assume that the error distribution is Gaussian in shape and to multiply the quoted error bars by 0.61 so as to produce one-sigma values.  Nevertheless, this practice still has to be remembered every time, and occasionally the writer (e.g., Krimm et al. 2009) does not tell the reader that 90\% error bars are used.

A complexity arises with many measured $E_{peak}$ values having asymmetric error bars, usually with the uncertainty towards high energy being much larger than the uncertainty towards lower energy.  This arises when $E_{peak}$ is near the upper end of the spectrum.  To illustrate this with an extreme example, consider a spectrum that shows a power law with a small amount of curvature up to a cutoff of 300 keV, in which case we can say that the $E_{peak}$ value is near 300 keV with a small uncertainty to low energies and an unlimited uncertainty to high energies.  This case arises frequently for the Swift satellite due to its fairly low energy cutoff.  The general solution is the tedious one of carrying asymmetric error bars for all quantities derived from the $E_{peak}$ values.

\section{Quantifying $\sigma_{Choice}$}

If two separate analysts independently report their $E_{peak}$ values, for the same burst as measured from the same satellite, with the exact same Poisson noise, using the same model, then the only difference is from the choices made by the analysts, $\sigma_{Choice}$.  Once a pairing of this kind is identified, for each burst the two analysts have in common, $\Delta$ is the logarithmic difference in the values the two analysts measured the burst.  The result of this will be a list of $\Delta$ values for the comparison pairs. From here, a simple calculation of the standard deviation of the $\Delta$ values will equal $\sigma_\Delta$. This scatter of $\Delta$ arises from the differences in the two individual sets of choices, so the uncertainty due to a single set of choices would simply be given by:
\begin{equation}
\label{eq:sigchoice}
\sigma_{Choice} = \frac{\sigma_\Delta}{\sqrt{2}}.
\end{equation}
Our procedure is to find published analyses which report $E_{peak}$ values for many identical bursts all using the exact same data from some satellite, to calculate a list of $\Delta$ values, and finally to calculate $\sigma_{Choice}$ from Equation 3.  With this, the ordinary variations in $E_{peak}$ caused by analysis choices will be attributed equally between the two analysts.

For the \textit{BATSE} era, we can compare values from Band et al. (1993), Yonetoku et al. (2004), Kaneko et al. (2006), and Nava et al. (2008).  For example, \textit{BATSE} trigger 1025 has reported $E_{peak}$ values of 114, 153, 131, and 117 keV for the four sources, while trigger 451 has 40, 134, and 143 keV for the first three sources respectively.  For the \textit{BATSE} era, we present the results in Table \ref{ChoiceBATSE}. The Kaneko-Yonetoku pair has the lowest scatter, which is about half that of the Band-Kaneko pairing and about a quarter of that of the Band-Nava pairing. This indicates that the choices made by Kaneko and Yonetoku are typically more alike than the choices made by any other pair.  We cannot identify any one analyst as producing better results.

For the \textit{Swift} era, we have just the one pairing to consider, the published values of Sakamoto et al. (2008) and Butler et al. (2007). Here, we use the Butler values of $E_{peak}$ that were derived from frequentist statistics, as that is what Sakamoto used in obtaining his values. We find 23 common bursts to use in this pairing, resulting in $\sigma_{Choice} = 0.04$.  We speculate that the reason for this Swift $\sigma_{Choice}$ being much smaller than the BATSE values (see Table 1) is that the coded mask of Swift eliminates the uncertainties in the background subtraction.

To further illustrate the effects of sigma choice, we present two figures (Figures 1 and 2) to display two of our comparison sets with the BATSE data. Figure 1 plots $E_{peak}$ from Yonetoku et al. (2004) vs. Kaneko et al. (2006). Figure 2 plots $E_{peak}$ from Yonetoku et al. (2004) vs. Nava et al. (2008). In both cases, we represent the bursts with a diamond with their associated error bars. We also plot a solid line in each of these figures to represent where the bursts should lie in an ideal world (i.e. in total agreement). As we described earlier, in these data pairs the only difference in the analysis is the choices made by the analysts.

There is a significant scatter on the value of $E_{peak}$ that can be attributed purely to the choices we make as analysts in deriving these values.  We now have six different values of $\sigma_{Choice}$; 0.07, 0.15, 0.21, 0.14, 0.29, and 0.04.  The $\sigma_{Choice}$ can vary by up to a factor of six.  For any analyst, we can only use the average.  A simple average is 0.15.  Likely, a better representation is the weighted average where the weights equal the number of bursts, for which we get $\sigma_{Choice}=0.12$.  This typical value of $\sigma_{Choice}$ is daunting in size.  For an example with $E_{peak}=100$ keV, the one-sigma range (from $\sigma_{Choice}$ alone) would be from 76-132 keV, which is nearly a factor of two in total size.

\section{Measuring $\sigma_{Det}$}

$\sigma_{Det}$ is the uncertainty associated with three particular problems related to the detector response. The first of these issues is associated with imperfect knowledge of the detector response.  The second issue is how the energy ranges of various detectors are different and thus could yield different values for $E_{peak}$. A third issue lies in the detector thresholds in that bursts for which the peak energy is just above the detector threshold will have ill defined spectral indices and therefore will not be well measured. In principle, this can be measured by comparing $E_{peak}$ values for measures of individual bursts with different detectors.  Care must be taken that these compared values were made over the whole time interval of the burst and with an identical model.  The procedure is to tabulate $\Delta$ values for many bursts observed with pairs of satellites, with the RMS scatter of $\Delta$ being related to $\sigma_{Det}$.  The statistical error bars ($\sigma_{Poisson}$) for each measure are known and can be accounted for.  In principle however, the effects of $\sigma_{Choice}$ and $\sigma_{Det}$ cannot be separated out.  So what we can take from this comparison of $E_{peak}$ values from different detectors is just the combined uncertainty, $\sigma_{Sat} = \sqrt{\sigma_{Choice}^2 + \sigma_{Det}^2}$.

The uncertainty in each $\Delta$ comes from the statistical uncertainty for each satellite and the $\sigma_{Sat}$ for each satellite;
\begin{equation}
\label{eq:sigdrm}
\sigma^2_\Delta = \sigma^2_{Poisson,1} + \sigma^2_{Poisson,2} + \sigma^2_{Sat,1} + \sigma^2_{Sat,2}
\end{equation} 
Where the numbers in the subscripts identify the two satellites.  In practice, we cannot distinguish the separate systematic effects of the two detectors, so all we can do is take $\sigma_{Sat}$ as the average of the two satellites.

The $\Delta$ values will be for bursts with a wide range of statistical errors, with each individual value being a Gaussian distribution with standard deviation of $\sigma_{\Delta}=\sqrt{ \sigma^2_{Poisson,1} + \sigma^2_{Poisson,2} +2 \sigma^2_{Sat}}$.  So the quantity $\Delta / \sigma_{\Delta}$ should be distributed as a Gaussian with a standard deviation of unity.  Our procedure is to vary $\sigma_{Sat}$ until the RMS scatter of $\Delta / \sigma_{\Delta}$ equals 1.

We have collected published $E_{peak}$ values for many bursts as measured by many satellites.  We then identified pairs of measures for individual GRBs that have identical models and that cover the entire time interval of the burst.  For each pair of satellites, we then calculate the $\sigma_{Sat}$ such that the $\Delta / \sigma_{\Delta}$ values have an RMS of unity.  These values are given in Table 2.  Two effects should be considered when viewing these results.  First, the entire second column, involving Swift-Suzaku and either Swift or Suzaku, involves joint data in the comparison, so the differences in the two $E_{peak}$ values will be smaller than if the two spectra were totally independent.  Thus we will not use these two values in evaluating an overall average $\sigma_{Sat}$.  Second, whenever a small number of bursts are involved, random fluctuations in $E_{peak}$ will lead to large variations in $\sigma_{Sat}$.  To take an extreme example, if only one burst is considered and the two measures are randomly close together, then the $\sigma_{Sat}$ value will be near zero.  Indeed, we see for the entire right-hand column of Table 2, with all entries coming from 2-4 GRBs, that all entries are at the extremes of the range.  These three entries have a total of 9 comparisons, and we combine them to form a single $\sigma_{Sat}$ involving RHESSI versus other satellites, with this value being 0.14.

So we now have a number of measures for $\sigma_{Sat}$, one from the Suzaku column, three from the Konus-Wind column, and one combined value for RHESSI.  These values range over a factor of two, from 0.08 to 0.16.  A straight average of these five measures is 0.13.  A weighted average involving the number of bursts in each measure yields 0.12.  We take this last value to be characteristic and average for a wide range of detectors and analysts.

We have now concluded that the global average $\sigma_{Sat}=0.12$ and $\sigma_{Choice}=0.12$.  Formally, this implies that $\sigma_{Det}=0$, but we can really only conclude that $\sigma_{Det}$ is negligibly small.  This provides confidence that the detector calibrations are well done.  In other words, the systematic differences from satellite to satellite are negligible, whereas the often-large differences from satellite to satellite are apparently caused simply by the ordinary choices made by the individual analysts.

\section{Measuring $\sigma_{Def}$}

Previous workers have defined $E_{peak}$ in a variety of different ways, with the resultant variations leading to an uncertainty labeled $\sigma_{Def}$.  This definitional uncertainty can be broken into four components: $\sigma_{Model}$ for whether the Band model or the Comptonized power law model is adopted, $\sigma_{Peak?}$ for whether the spectrum is extracted for the entire burst or just the time interval around the peak flux, $\sigma_{Fixed \alpha \beta}$ for whether the analyst systematically fixes the high-energy and low-energy power law slopes in the Band function ($\alpha$ and $\beta$ respectively) to some average value, and $\sigma_{F/B}$ for whether the analyst uses frequentist fitting or uses Bayesian analysis with some set of adopted priors.  For each of these, we take the same approach we did with $\sigma_{Choice}$ and divide $\sigma_\Delta$ by a factor of $\sqrt{2}$. The overall uncertainty from these definition issues ($\sigma_{Def}$) will be just the addition in quadrature of the four components as applicable for the question in hand.

The definition of $E_{peak}$ (i.e., the photon energy for the maximum of $\nu F_{\nu}$) requires a fit to the spectrum, but it has not specified the functional form for this fit.  Most published values are roughly evenly divided between the Band function or the Comptonized power law (CPL).  There is a systematic offset in how $E_{peak}$ is measured in that the CPL model consistently predicts a higher $E_{peak}$ than the Band model (see Figure 6 of Krimm et al. 2009). This offset is expected, because the CPL falls off much faster than the Band function at high energies, so the CPL fit must push $E_{peak}$ to higher energies to match the observed spectra.  To measure this difference in $E_{peak}$ for a typical ensemble of GRBs, we have used the results from Krimm et al. (2009), where the Swift-plus-Suzaku spectra are fitted to both the Band function and the CPL.  In all, we can calculate $\Delta$ values for 67 bursts.  The average $\Delta$ is 0.14, while $\sigma_{Model}=0.12$.

The definition of $E_{peak}$ does not state the time interval over which the spectrum is to be extracted.  Indeed, the $E_{peak}$ values change fast throughout the entire burst, so there is a big problem in knowing what interval to use.  A unique solution is to take the entire burst.  This has the advantage of getting the best signal-to-noise ratio for the spectrum (unless the burst is not sufficiently above the background).  A spectrum from the entire burst (the fluence spectrum) makes logical sense for use with the Amati and Ghirlanda relations, both of which connect with the burst fluence.  An alternative solution is to use the $E_{peak}$ value for the time interval around the time of the peak flux in the burst light curve.  This solution is logical for all the other luminosity relations that connect with the burst peak luminosity, as then both the $E_{peak}$ and luminosity will correspond to the same time and physics.  An ambiguity arises in specifying the duration of the interval, where this interval might be constant, scale with the (perhaps unknown) redshift, or scale with the burst or pulse duration.  The point is that alternative solutions will lead to a systematic variation in $E_{peak}$, and this uncertainty will be labeled as $\sigma_{Peak?}$.  To evaluate this, we have used the many fits reported by Krimm et al. (2009) for 28 GRBs as measured by Swift-plus-Suzaku.  For these bursts, they report $E_{peak}$ for the Band function for both the entire burst as well as a tight interval centered on the peak in the light curve, and for these we have calculated the $\Delta$ and total $\sigma_{Poisson}$ values.  As in Section 6, we calculate $\sigma_{Peak?}$=0.06.

The definition of $E_{peak}$ can use the Band model with or without fixed high-energy and low-energy power law slopes, and this change of definition will lead to a variation labeled as $\sigma_{Fixed \alpha \beta}$.  In general, the Band function is fitted with both $\alpha$ and $\beta$ as free parameters.  However, in practice, spectra rarely extend far past $E_{peak}$ which provides little constraint on $\beta$. This is often solved by simply setting the power law slope equal to some average value.  The original paper on the Band function (Band et al. 1993) provides 53 bursts with alternative fits where the slopes are allowed to vary freely or are fixed at $\alpha=-1$ and $\beta=-2$.  For these bursts, we have calculated the values of $\Delta$.  The average $\Delta$ is -0.07 while the RMS scatter is 0.21.  This average is marginally different from zero in the sense that the fixed-slope values are larger than the values with freely-fitted-slopes.  In all, $\sigma_{Fixed \alpha \beta}=0.15$.

The usual definition of $E_{peak}$ relies on frequentist methods (i.e, chi-square minimization of spectral models), whereas another possibility is to use Bayesian methods.  The Bayesian approach explicitly assumes sets of priors, where each prior quantifies the likely distribution of values.  This Bayesian method has been used in only one paper (Butler et al. 2007), and unfortunately, this paper made a variety of poor assumptions for the priors.  Most importantly, they assumed that the probability of the $E_{peak}$ values above 300 keV falls off fast as a log-normal distribution, and this means that the bursts with high $E_{peak}$ values will have their values pushed to greatly lower energy.  The fallacy of this assumption is demonstrated by a comparison of their $E_{peak}$ values with those from Suzaku, Konus-Wind and RHESSI.  For example, GRB 051008 has a measure of $E_{peak}=266^{+349}_{-80}$ keV from Butler et al. (2007), while Konus-Wind reports $865^{+107}_{-81}$ keV (Golenetskii et al. 2005b), Suzaku reports $1167^{+1078}_{-427}$ keV (Ohno et al. 2005), and Swift-plus-Suzaku reports $815^{+54}_{-47}$ keV (Krimm et al. 2009).  Of the 11 Konus-Wind GRBs with $E_{peak}>600$ keV, all 11 Butler et al. (2007) values are smaller (whereas only half should be smaller if the Bayesian prior was reasonable), with typical errors of a factor of 2.  Another mistaken prior is that they assume the $\beta$ values to follow a simple exponential distribution, with the result being to push the $\beta$ values greatly negative (making for a claimed high-energy cutoff that is too sharp hence pushing $E_{peak}$ to larger values).  Butler et al. (2007) give fitted $E_{peak}$ values by both frequentist and Bayesian methods for the exact same data for many bursts, and for each of these we have calculated the $\Delta$.  For 75 bursts for which the frequentist methods return a value instead of a limit (i.e., the case where the troubles with the priors are minimized), we find that the RMS of $\Delta$ is 0.07, and this is the value of $\sigma_{F/B}$.

We now have measures of $\sigma_{Model}=0.12$, $\sigma_{Peak?}=0.06$, $\sigma_{Fixed \alpha \beta}=0.15$, and $\sigma_{F/B}=0.07$.  In a situation where all four uncertainties are operating fully, the total uncertainty caused by the variations in the definition would be the sum in quadrature of the four components, with $\sigma_{Def}=0.21$.  This would correspond to a one-sigma uncertainty of a factor of 1.62.

Which of these uncertainties are applicable depends critically on the situation.  Here are four typical situations, each with different answers:  (1) If we are trying to compare an observed $E_{peak}$ value with some measure of a particle energy distribution, then it is completely unclear how to connect the two, so a full $\sigma_{Def}=0.21$ is appropriate.  That is, the Band function is a completely empirical description of the turnover in the spectrum, so it is unknown what part of the spectrum corresponds with any point in a calculated theoretical particle distribution.  (2) If a Hubble diagram is constructed using luminosity relations where the calibration and bursts all use exactly the same definition, then $\sigma_{Def}=0$.  This might be the case if all the $E_{peak}$ values are pulled from a single paper, or the case if we are anticipating some future program designed for the purpose.  (3) If the luminosity function is calibrated with a particular definition but then applied to a set of $E_{peak}$ values with a mixed set of definitions, then the contribution will be only a fraction of its full value.  For a data set that involves a fraction `$f$' of values made with the alternative definition, the $\sigma$ value will be $\sqrt{f}$ times the full value.  For example, the BATSE $E_{peak}$ values presented in Nava et al. (2008) have $f=0.31$ of the bursts with fixed $\alpha$ or $\beta$, so we would have $\sigma_{Def}=\sqrt{0.31}\sigma_{Fixed\alpha \beta} \approx 0.08$.  (4) If the Amati relation is evaluated with bursts from a wide array of detectors, then the mixed sets of definitions will lead to a partial contributions from the various alternative definitions used.  Schaefer (2007) has calibrated the Ghirlanda and $E_{peak}-L$ relations with bursts from BATSE, BeppoSAX, Konus, INTEGRAL, and Swift, and we estimate that $\sigma_{Def}\approx 0.15$.

The contributions to $\sigma_{Def}$ change greatly with the question being asked.  The contributions will also change substantially with the data set being used.  Not only will the fractions `$f$' change, but the size of the unmixed contribution will change.  For example, $\sigma_{F/B}$ will change greatly with the adopted priors, while $\sigma_{Fixed \alpha \beta}$ will change greatly depending on the adopted power law slopes.  In practice, it is impossible to evaluate meaningful error bars for the various contributions, because they change for every circumstance.  Therefore, the quantitative measures of the contributions to $\sigma_{Def}$ in this section can only be taken as approximate or maybe as typical, and each of the definitional alternatives leads to variation with an RMS scatter of roughly 0.1-0.2 (i.e., 23\% to 46\% errors).  Depending on the situation, the resulting $\sigma_{Def}$ might vary anywhere from 0.0-0.2.

\section{Are GRBs Thermostated?}

The distribution of $E_{peak}$ for GRBs has been observed to be fairly narrow (Mallozzi et al 1995), and thus the intrinsic scatter of $E_{peak}$ must be narrow as well. We can relate the observed and intrinsic values of $E_{peak}$ by a simple equation:
\begin{equation}
E_{peak,Obs} = E_{peak,Int}\left(1+\textit{z}\right)^{-1} \eta
\end{equation}
The observed $E_{peak}$ is related to the intrinsic $E_{peak}$ by the cosmological redshift factor of $(1+\textit{z})$. The factor $\eta$ encompasses all the various effects that lend to the imperfect measure of $E_{peak}$, with the RMS scatter of $\log \eta$ equaling $\sigma_{Total}$. Since these factors are multiplicative, it is more appropriate to evaluate this equation in log space. Therefore, the expression for the distribution of $E_{peak}$ in log space can be given as:
\begin{equation}
\sigma^2_{\log Ep,Obs} = \sigma^2_{\log Ep,Int} + \sigma^2_{\log(1+z)} + \sigma^2_{Total}
\end{equation}
We can readily quantify many of these values from data in hand. The \textit{Swift} website provides a list of confirmed spectroscopic redshifts from which we can compose a list of $\log_{10}(1+z)$, and find that RMS scatter is 0.19, which we can reasonably use as a typical value of $\sigma_{\log (1+z)}$. Likewise, we can use published data sets to get an estimate for $\sigma_{\log Ep,Obs}$. We can find the scatter in the log of the observed $E_{peak}$ from Brainerd et al. (1999) (this is a different GRB sample from the sample used to find $\sigma_{\log (1+z)}$). In this paper, the authors found the full width half-maximum of the BATSE $E_{peak}$ distribution to be 0.796 in log of the $E_{peak}$, which equates to a one-sigma scatter of $\sim 0.34$, which we use as the value of $\sigma E_{peak,Obs}$. Putting all these values together, we get:
\begin{equation}
\sigma^2_{\log Ep,Int} = 0.08 - \sigma^2_{Total}
\end{equation}
We see that $\sigma^2_{Total}$ needs to be 0.08 in order for the intrinsic scatter of $E_{peak}$ to be zero. This equates to a $\sigma_{Total} \sim 0.28$.

In previous sections, we have identified what goes into $\sigma_{Total}$, so we can easily calculate how our expected values for $\sigma_{Total}$ compare to what kinds of $\sigma_{Total}$ needed if the width of the distribution of $E_{peak}$ in the burst rest frame is `zero'. We find that $\sigma_{Poisson}$ has typical values near 0.15 for collections of bursts (with the values for individual bursts varying greatly with the detector and the burst brightness), with an extreme range of 0.04 to 0.35.  We find the average $\sigma_{Choice}=0.12$ and $\sigma_{Sat}=0.12$ (so that $\sigma_{Det}$ is near zero), with extreme values of 0.08 and 0.16.  We find that $\sigma_{Def}$ depends critically on the application, but typical applications might have values of 0.15, with extremes of 0.0 to 0.2.  These sources of error are independent, so they should be added in quadrature.  For these typical values, the $\sigma_{Total}=0.24$, with an extreme range of 0.09 to 0.43. So the $\sigma_{Total}$ needed for $\sigma_{\log Ep,Int}$ to be zero is not only within our expected range of $\sigma_{Total}$, it is a typical value for $\sigma_{Total}$. The value of $\sigma_{\log Ep,Int}$ will be small for any realistic value of $\sigma_{Total}$. Even if one were to take a the lowest estimate of $\sigma_{Total} = 0.09$, we would still have a small value of $\sigma_{\log Ep,Int}=0.28$. This means that for the observed distribution of $E_{peak}$ to be as narrow as observed, the intrinsic rest frame distribution of $E_{peak}$ must also be narrow in all cases.

In order to be sure that our choice of $E_{peak,Obs}$ is appropriate, we must be certain that selection effects are not causing a perceived distribution. An example of this is in Sakamoto et al. (2008), where there are clear cutoffs for different instruments depending on the energy range of a detector's energy threshold. This is why we exclusively use the BATSE data to determine this value. In Brainerd et al. (1999), the authors found that the BATSE trigger thresholds did not cause the observed distributions. The X-Ray flashes (e.g. Sakamoto et al., 2005; P\'{e}langeon et al., 2008) are just the tail of the observed classical burst distribution, or at most a small excess out on the tail. Another important part of the findings of Brainerd et al. (1999) is that the detector thresholds are not causing an artificial distribution in the detected bursts. Figure 3 of Brainerd et al. (1999) shows a simulated histogram for the detection of bursts for a given power-law distribution of $E_{peak}$. The results show that the distribution the detection of $E_{peak}$ has roughly the same efficiency on either side of the distribution. This implies that the narrowness of the BATSE distribution is not being artificially cut off by some sort of systematic effects on the part of the detector threshold. It is for this reason that we believe that the BATSE data at the very least shows a real distribution for $E_{peak}$, not an artifact of selection effects. Therefore, our finding of no scatter in the intrinsic $E_{peak}$ distribution is sound.

Another method for showing that the distribution of $E_{peak,Int}$ is small is to use a large sample of data for which there are known bursts across a wide range of known redshifts. Using the known redshifts, we can find the scatter of $E_{peak,Obs}(1+\textit{z})$ directly. In doing this, we can, in a sense, eliminate one of our terms in the earlier method. The uncertainty we find by finding the standard deviation of $E_{peak,Obs}(1+\textit{z})$ will instead be quantified as:
\begin{equation}
\sigma^2_{E_{peak,Obs}(1+\textit{z})} = \sigma^2_{E_{peak,Int}} + \sigma^2_{Total}.
\end{equation}
For this purpose, we can use the large data set available in Schaefer (2007). We find the RMS scatter of $E_{peak,Obs}(1+\textit{z})$ of the whole data set to be 0.47. If we were to adopt a typical value of $\sigma_{Total} = 0.30$, we get $\sigma_{E_{peak,Int}} = 0.37$, which is still a fairly narrow distribution. While it is not as small as the first test showed, it is nonetheless narrow, showing that the one sigma scatter of $E_{peak,Int}$ is merely a factor of  $\sim2$. 

We should also address the possibility of mixing bursts with widely different redshifts in our sample. To do this, we bin up the Schaefer data by redshift ranges 0-1, 1-2, 2-3, and 3-5. We find the scatter of $E_{peak,Obs}(1+\textit{z})$ for these bins to be 0.62, 0.40, 0.35 and 0.27 respectively. In addition, we have applied a similar test to the \textit{Swift}-Suzaku data (Krimm et al. 2009). For this data, we bin the bursts by redshifts 0-1, 1-2, and 2-4. We find the RMS scatter of  $E_{peak,Obs}(1+\textit{z})$ for these bins to be 0.42, 0.44, 0.24 respectively. With seven different bins, we can see that the median value is 0.40. Using the Schaefer (2007) data, we find that the average value of the log of $E_{peak,Obs}(1+\textit{z})$ is 2.23, 2.58, 2.52 and 2.60 for their respective bins, and for the Krimm et al. (2009) data we find the average values to be 2.66, 3.03, 2.89 respectively. Therefore, there are no visible trends with redshift. Indeed, this shows that the average value of $E_{peak,Obs}(1+\textit{z})$ is close to 511 keV. We believe that the narrowness of $E_{peak}$ is therefore physical and not the result of selection effects.

With this finding, the obvious question is what is the mechanism driving \textit{all GRBs} to have the same (or essentially the same) intrinsic $E_{peak}$. With the rest frame $E_{peak}$ values being like the effective temperature of the gamma-ray emitting region, the nearly constant temperature requires some mechanism to act as a thermostat, holding the temperature at a fixed value.  Our realization that the rest frame $E_{peak}$ is nearly a constant is new, with this conclusion being simple and forced. The task for our community is now to understand the physical mechanism for this thermostat effect.

The typical values of $E_{peak,Obs}(1+\textit{z})$ is nearly comparable to the electron rest-mass energy $(m_{e} c^2)$ of 511 keV. This suggests that the thermostat mechanism involves an equilibrium between electron-positron pair creation and annihilation. 

\section{Implications}

We have now identified the various sources of scatter on $E_{peak}$. We have found $\sigma_{Poisson}$ to have typical values near 0.15, although this has a large range of 0.04 to 0.35. The reason for this range lies mostly due to the detector and the brightness of the burst. We find that typically $\sigma_{Choice}$ is on the same order of $\sigma_{Sat}$, 0.12. This indicates that the scatter due to the detector itself, $\sigma_{Det}$ is small. We find an extreme range of 0.08 and 0.16 for the error due to analyst choices. The uncertainty associated with the definition of $E_{peak}$ depends on the application and a range of 0.0 to 0.2 for $\sigma_{Def}$, with a typical value of 0.15. Finally, if these are all put together, we find that $\sigma_{Total}$ to have a range of 0.09 to 0.43. For the typical values, we find that we should expect that $\sigma_{Total}=0.24$.

One important implication is that there is a real limit on the accuracy with which any $E_{peak}$ can be measured.  Even for a very bright burst with a well placed $E_{peak}$, say with $\sigma_{Poisson}=0.01$, and some agreed-upon definition (so $\sigma_{Def}=0$), we still always have $\sigma_{Total}=\sigma_{Choice}=0.12$.  We know of no realistic way to measure or legislate or even define the `best' choices by analysts, so this limit cannot be improved.  This means that all GRBs must have at least a 28\% error in $E_{peak}$.

A second implication that can come from this result is that $E_{peak}$ has accuracy limits as a  luminosity indicator. On an individual basis, this is a valid limitation.  Supernovae have a similar limitation, although their real systematic uncertainty for an individual event is 2.1$\times$ better than for GRBs. We can overcome the accuracy limitation in the same way as for supernovae by using large numbers of bursts. Therefore, we can bring down the uncertainty by a factor of the square root of the number of bursts.  Again, while Gamma-Ray Bursts have less accuracy than supernovae, they make up for it in their unique coverage at high redshifts.

A third important implication is that collections of bursts have a greatly larger average error than is realized in our community.  All collections of $E_{peak}$ values have mixed definitions and few bright bursts, so $\sigma_{Total} \approx 0.24$ is the norm.  This corresponds to a 55\% error.  For a burst claimed to be $E_{peak}=100$ keV, the real total 1-sigma error region will be like 58-174 keV, regardless of the published statistical error bar. 

An important implication of this work is that it implies that GRBs have their emission region effectively held to a constant temperature by some thermostat mechanism. That is the observed $E_{peak}$ distribution is already fairly narrow, so the intrinsic distribution of $E_{peak}$ in the burst rest frame must be very narrow. We have no conclusive mechanism to cause this, but one such explanation is that electron-positron annihilation may be acting as a thermostat for GRB emission.

A final important implication is to that our community can improve the measurement of $E_{peak}$ for many purposes.  We cannot think of any realistic or effective way to legislate the analyst's choices.  But we can make sure that all the $E_{peak}$ values being used have only one definition.  This will require a uniform analysis, which might be accomplished by having one analyst processing all the bursts used in the sample.  Or it might require that multiple analysts agree to adopt some standard definition.  For this, we suggest the standard be based on the Band function with freely varying $\alpha$ and $\beta$ and a frequentist chi-square minimization for the entire burst time interval.

{}

\begin{deluxetable}{llll}
\tablewidth{0pc}
\tabletypesize{\small}
\tablecaption{$\sigma_{Choice}$ in the \textit{BATSE} Era \tablenotemark{a}}
\tablehead{\colhead{} & \colhead{Yonetoku\tablenotemark{c}} & \colhead{Band\tablenotemark{d}} & \colhead{Nava\tablenotemark{e}}}
\startdata
Kaneko\tablenotemark{b}    &  0.07 (75) &  0.15 (11)    &  ---      \\
Yonetoku\tablenotemark{c}   & ---              &  0.21 (34)    &  0.14 (62)          \\
Band\tablenotemark{d}   & ---              &  ---      &  0.29 (5)        \\
\enddata
\tablenotetext{a}{The values reported in this table are $\sigma_{Choice}$, which are the uncertainties in $\log_{10}E_{peak}$ due to the particular choices made by {\it one} analyst.  The following number in parentheses is the number of common bursts that were used for the calculation.}
\tablenotetext{b}{Kaneko et al. (2006)}
\tablenotetext{c}{Yonetoku et al. (2004)}
\tablenotetext{d}{Band et al. (1993)}
\tablenotetext{e}{Nava et al. (2008)}
\label{ChoiceBATSE}
\end{deluxetable}

\begin{deluxetable}{lllll}
\tablewidth{0pc}
\tabletypesize{\small}
\tablecaption{$\sigma_{Sat}$ in the \textit{Swift} Era\tablenotemark{a}}
\tablehead{\colhead{} & \colhead{Suzaku\tablenotemark{c}} & \colhead{\textit{Swift}-Suzaku\tablenotemark{d}}  & \colhead{Konus-Wind\tablenotemark{c}} & \colhead{RHESSI\tablenotemark{c}}}
\startdata
\textit{Swift}\tablenotemark{b}                  & 0.15 (7)               & 0.04\tablenotemark{e}  (8)               & 0.16 (13)    & ---  \\
Suzaku\tablenotemark{c}                          & ---                        & 0.03\tablenotemark{e}  (11) 	     & 0.08 (23)  & 0.18 (3)\tablenotemark{f} \\
\textit{Swift}-Suzaku\tablenotemark{d}    &  ---                       & ---                         & 0.12 (23)  & 0.02 (2)\tablenotemark{f}  \\
Konus-Wind\tablenotemark{c}                 & ---                        & ---                          & ---                & 0.18 (4)\tablenotemark{f} \\
\enddata
\tablenotetext{a}{The values reported in this table are $\sigma_{Sat}$, which are the one-sigma uncertainty of $\log_{10}(E_{peak})$ for the combined causes of uncertainties in one detector response and one analyst's choices.  The following number in parentheses is the number of common bursts that were used for the calculation.} 
\tablenotetext{b}{Sakamoto et al. (2008) and Butler et al. (2010)}
\tablenotetext{c}{Multiple GCNs}
\tablenotetext{d}{Krimm et al. (2009)}
\tablenotetext{e}{This entry is a comparison between composite spectra from Swift-plus-Suzaku versus spectra from one part of that composite.  So the resulting $E_{peak}$ values are not independent, as the joint part will share identical data, identical Poisson noise, and identical detector response measures.  As such, the $\Delta$ values will be systematically smaller, and hence our $\sigma_{Sat}$ will be smaller than expected for the case where the input data was completely independent.}
\tablenotetext{f}{RHESSI comparisons have small number statistics, so we combine the three measurements in a weighted average to get one singular measurement of $\sigma_{Sat}$ = 0.14 (9).}
\label{DRMSWIFT}
\end{deluxetable}

\clearpage

\begin{figure}
\begin{center}
\includegraphics[scale=0.8]{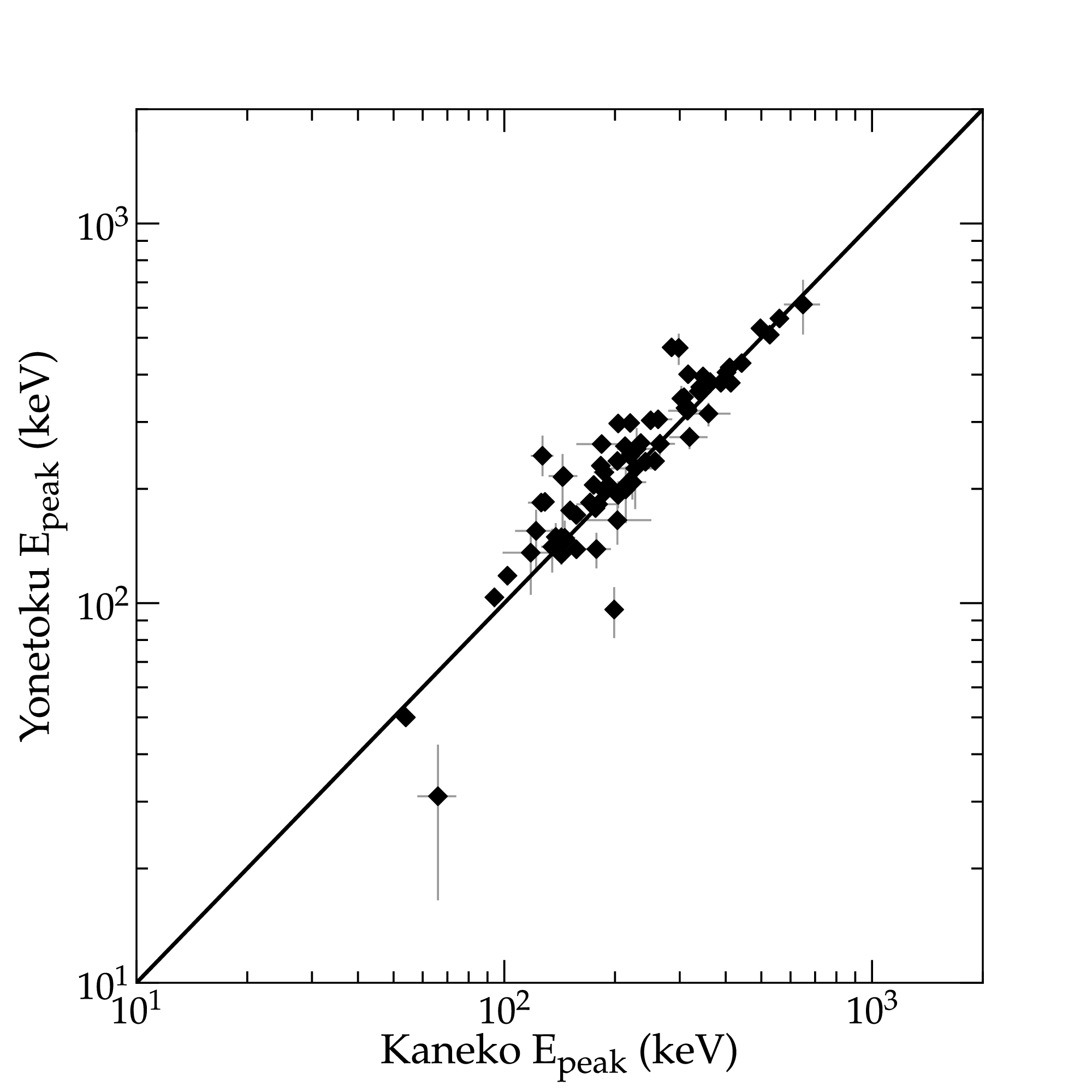}
\end{center}
\caption{A visualization of the scatter in $E_{peak}$ due to $\sigma_{CHOICE}$ alone. Here we compare BATSE bursts (diamonds) as measured by two groups of analysts - Yonetoku et al. (2004) vs. Kaneko et al. (2006). The solid line denotes the ideal case where both groups would be in complete agreement. The scatter about the diagonal line is $\sigma_{Choice}$, and the point of this figure is that there is significant scatter even for identical bursts, identical Poisson noise, and identical data.}
\label{fig:Kanektoku}
\end{figure}

\clearpage

\begin{figure}
\begin{center}
\includegraphics[scale=0.8]{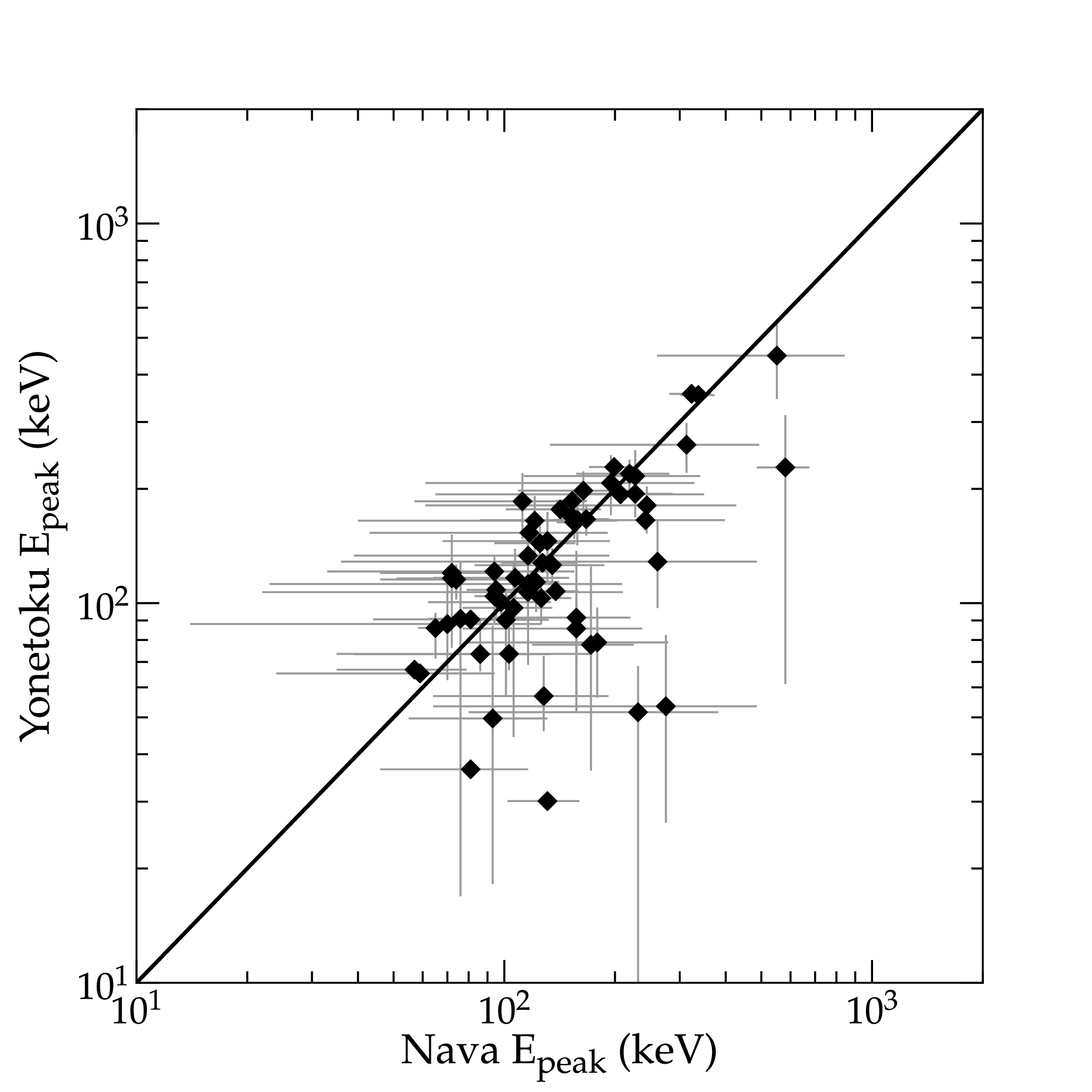}
\end{center}
\caption{A visualization of the scatter in $E_{peak}$ due to $\sigma_{CHOICE}$ alone. Here we compare BATSE bursts (diamonds) as measured by two groups of analysts - Yonetoku et al. (2004) vs. Nava et al. (2008). The solid line denotes the ideal case where both groups would be in complete agreement. The two analysts compared identical bursts, with identical data, and with identical models; so the large scatter about the diagonal proves that individual unrecorded choices by the analysts have a large effect on the reported $E_{peak}$.}
\label{fig:Navatoku}
\end{figure}

\end{document}